# One-dimensionalization by Geometrical Frustration in the Anisotropic Triangular Lattice of the 5$d$ Quantum Antiferromagnet Ca$_3$ReO$_5$Cl$_2$

Daigorou Hirai[1]*, Kazuhiro Nawa[1], Mitsuaki Kawamura[1], Takahiro Misawa[1], and Zenji Hiroi[1]

[1] *Institute for Solid State Physics, University of Tokyo, Kashiwa, Chiba 277-8581, Japan*
*e-mail address: dhirai@issp.u-tokyo.ac.jp*

We report on the emergence of antiferromagnetic spin chains from two-dimensionally aligned spins on the anisotropic triangular lattice (ATL) in the insulating calcium rhenium oxychloride Ca$_3$ReO$_5$Cl$_2$. The compound contains Re$^{6+}$ ions each with one unpaired electron in the $d_{xy}$ orbital, which are arranged to form a spin-1/2 ATL with $J'/J \sim 0.32$ and $J = 41$ K, where $J$ and $J'$ are magnetic interactions in and between the chains, respectively. In spite of the apparent two dimensionality, we observe clear evidence of a gapless spin liquid that is characteristic of the spin-1/2 Heisenberg chain. This one-dimensionalization must be caused by geometrical frustration: growing antiferromagnetic correlations in every chain effectively cancel out inter-chain zigzag couplings at low temperature. Ca$_3$ReO$_5$Cl$_2$ provides us with a detailed insight into the interesting physics of the ATL antiferromagnet, especially via comparison with the typical ATL compound Cs$_2$CuCl$_4$.

## I. INTRODUCTION

Dimensionality is one of the most fundamental factors that determine the magnetism of materials [1–3]. For the one-dimensional (1D) spin-1/2 Heisenberg antiferromagnetic (AF) chain, a Tomonaga–Luttinger liquid (TLL) is realized [4], which is characterized by gapless spin-wave-like excitations at wave vectors $q = 0$ and $\pi$ and continuous spin excitations above those, known as the two-spinon continuum [5,6]. These gapless excitations give rise to a finite magnetic susceptibility at $T = 0$ and a $T$-linear heat capacity at low temperatures [7,8].

There are many compounds studied thus far as candidates for low-dimensional magnets. They contain low-dimensional networks made of magnetic ions embedded into three-dimensional (3D) crystal structures. For 1D magnets, for example, Sr$_2$CuO$_3$ is a typical compound for the spin-1/2 1D Heisenberg antiferromagnet (HAFM), which comprises vertex-sharing CuO$_4$ chains with the intra-chain magnetic interaction of $J \sim 2,000$ K and a small inter-chain interaction of $J'/J \sim 10^{-5}$ [9,10]. It exhibits clear 1D characters at finite temperatures; a broad peak in magnetic susceptibility that is reproduced by the so-called Bonner–Fisher curve and a $T$-liner heat capacity scaled by $J$. However, it undergoes an AF LRO below 5 K owing to small inter-chain interactions; such behavior is generally observed in quasi-low-dimensional AF compounds.

On the other hand, several compounds exhibit 1D magnetism although their crystal structures do not have apparent one dimensionality. For example, KCuF$_3$ crystallizes in a 3D perovskite structure, but the dominant magnetic interaction occurs only along the $c$ axis. This magnetic one dimensionality originates from the specific arrangement of the $d_{x2-y2}$ orbitals carrying spin-1/2: the $d_{x2-y2}$ orbitals form $\sigma$ bonding via oxide ions along the $c$ axis to give a large AF coupling, while they are orthogonal to each other in the $ab$ plane to give a small ferromagnetic coupling [11].

Another source of dimensional reduction is geometrical frustration. In the spin ladder compound SrCu$_2$O$_3$, for example, a couple of Cu chains or a Cu ladder along the $b$ axis is connected with neighbors along the $a$ axis by a halfway shift along the $b$ axis to form a two-dimensional (2D) network, in which the 1D arrays of squares and isosceles triangles alternate along the $a$ axis [12]. The magnetic interactions are largely antiferromagnetic, $\sim 3,000$ K, at the square, while weakly ferromagnetic, $\sim -100$ K, at the isosceles triangle [13]. The compound has a spin-gapped ground state expected for an isolated spin ladder in spite of the sizable inter-ladder zigzag coupling. In addition, another spin ladder compound Sr$_2$Cu$_3$O$_5$ with three legs exhibits a TLL [13]. The reason of one-dimensionalization in these compounds is ascribed to the geometrical frustration at the inter-ladder zigzag coupling: magnetic interactions are completely canceled out at the isosceles triangle only when AF correlations have developed largely enough along the ladder at low temperatures [14].

In order to study one-dimensionalization by geometrical frustration in a systematic way, we focus on the anisotropic triangular lattice (ATL) HAFM, which comprises two kinds of magnetic interactions, $J$ and $J'$, on a triangular lattice, as illustrated in Fig. 1. The ATL model spans a range between decoupled 1D chains in the $J'/J = 0$ limit and an isotropic 2D triangular lattice at $J'/J = 1$. Thus, one can study the effect of the geometrical frustration at the inter-chain zigzag coupling as a function of the $J'/J$ ratio. The ground state of the regular triangular lattice is well established as a three-sublattice noncollinear magnetic order where the three neighboring spins have a 120° spin rotation with each other (the so-called 120° order) [15–18]. Indeed, this type of magnetic order has been experimentally observed in a spin-1/2 regular triangular antiferromagnet [19]. In contrast, the ground state of the decoupled 1D chains is a TLL, as mentioned above [20,21].

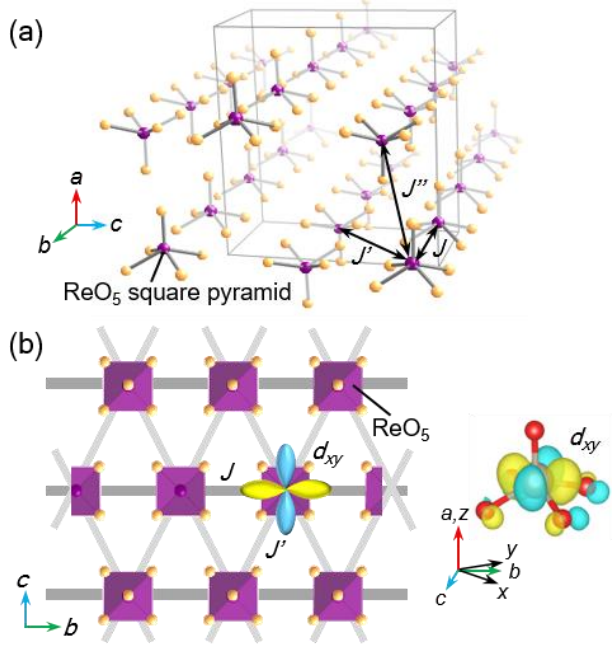

Fig. 1. (a) Arrangement of $ReO_5$ square pyramids in the orthorhombic crystal structure of $Ca_3ReO_5Cl_2$ [space group *Pnma*, $a$ = 11.892(3) Å, $b$ = 5.5606(12) Å, $c$ = 11.129(4) Å]. The Ca and Cl atoms are omitted for clarity. Three magnetic interactions ($J$, $J'$, and $J''$) between Re spins are shown by the arrows. (b) Anisotropic triangular lattice made of $Re^{6+}$ ions carrying spin-1/2 in the $d_{xy}$ orbital (blue and yellow robes on the Re ion) lying in the $bc$ plane. The nearest-neighbor magnetic interaction $J$ along the $b$ axis and the next-nearest-neighbor magnetic interaction $J'$ in the $bc$ plane are shown. A Wannier orbital occupied by a valence electron, which is dominantly composed of the $d_{xy}$ orbital, is also shown. The DFT calculations reveal the ratios of $J'/J$ = 0.295 and $J''/J$ = 0.0007.

The ground states of ATLs in the intermediate range (0 < $J'/J$ < 1) have not yet been completely understood by numerous calculations [22–37]. In the classical limit, a spiral magnetic order is expected in the whole range [22,23], while quantum fluctuations enhance one dimensionality particularly for small ratios [24,30,32–36]. As the result, a gapless spin liquid analogous to a TLL becomes stable in a wide range below 0.6–0.7; the critical ratios are scattered depending on the calculation methods [30,31,34,36,37]. As the ratio further increases, it is replaced by a dimer-ordered phase for 0.7 < $J'/J$ < 0.9 [30] or a gapped spin liquid phase for 0.6 < $J'/J$ < 0.8 [31,37], before a 120° LRO expected for the regular triangular lattice is realized near $J'/J$ = 1. As is always the case for the numerical calculations for frustrated spin systems, it may be cumbersome to choose one ground state out of several ones energetically competitive with each other. Therefore, a close collaboration between numerical studies and experiments on actual candidate materials are necessary for better understanding.

The ATL spin model is realized in a few classes of materials. $Cs_2CuCl_4$ is a typical compound with spin-1/2 $Cu^{2+}$ ions on an ATL. In fact, "dimensional reduction" observed in this compound [38,39] has triggered numerous studies on the ATL model. Despite the considerably large $J'/J \sim 0.3$ determined by various measurements [38–41], spinon continuum characteristic of a 1D spin system has been observed in inelastic neutron scattering (INS) measurements [38,39]. A recent calculation can reproduce the experimental data without any adjustable parameters [24,35], which clearly evidenced one-dimensionalization at this parameter range. On the other hand, model compounds are missing in the relatively large $J'/J$ region where expected are unknown quantum phases such as a dimer order or a gapped spin liquid phase. To clarify the ground states of ATL antiferromagnets, more model compounds that cover a wide range of the $J'/J$ ratio are highly desirable.

In the course of materials search in 5$d$ electron systems, which have been attracting great interest due to their unique physical properties in recent years [42,43], we found a new 5$d$ transition metal oxychloride $Ca_3ReO_5Cl_2$ (CROC for short) with $Re^{6+}$ ($5d^1$) ion [44]. It exhibits distinct pleochroism: the crystal changes its color depending on the viewing direction or the polarization of the incident light. This unique optical property is a consequence of the optical selection rule of the $d$–$d$ transition of the 5$d$ electron in the Re ion: the optical absorption occurs at the visible light range from the lowest $d_{xy}$ state to the higher $d$ levels in the square-pyramidal crystal field.

In the present study, we focus on the magnetic properties of CROC. The Re atom occupies only one crystallographic site, which is surrounded by five oxygen atoms in a slightly distorted square-pyramidal coordination [44]. The three-dimensional arrangement of the $ReO_5$ square pyramids is depicted in Fig. 1(a). The $ReO_5$ units are isolated from each other without sharing their oxygens, and half of them point upward along the [100] direction, and the rest point downward along the [−100] direction in a staggered manner along the $c$ axis. The neighboring magnetic interactions $J$, $J'$, and $J''$ are defined as shown in Fig. 1: there are two neighbors connected by $J$ along the $b$ axis, four $J'$ in the $bc$ plane which form zigzag couplings, and four by $J''$ out of the plane. The corresponding Re–Re distances are 5.5661(1), 6.3989(3), and 5.5515(3) Å, respectively. In spite of the seemingly 3D crystal structure, the magnetic lattice is approximated as an ATL spreading in the $bc$ plane owing to the specific arrangement of the $d_{xy}$ orbitals, as will be mentioned later.

Our magnetic susceptibility measurements demonstrate that $Ca_3ReO_5Cl_2$ is a good model compound for the spin-1/2 ATL HAFM with the $J'/J$ ratio of 0.32 and $J$ = 40.6 K. This experimentally obtained $J'/J$ ratio is comparable to the value of 0.295 estimated by the first-principles DFT calculations. In spite of the large $J'$ value, heat capacity measurements reveal a well-defined $T$-linear contribution with a large coefficient that is expected for a 1D HAFM with $J$ = 48.3 K. In addition, an LRO is observed at a much reduced temperature of $T_N$ = 1.13 K, which may be induced by the small inter-plane coupling $J''/J$ = 0.0007 as estimated by the first-principles calculations. The observed 1D character of CROC is likely due to the dimensional reduction by geometrical frustration and quantum fluctuations in the ATL, as observed for $Cs_2CuCl_4$ with $J'/J$ ~ 0.3 [38–41] and organic compounds [45–48]. The present study demonstrates that one-dimensionalization occurs ubiquitously over a wide variety of materials.

## II. EXPERIMENTAL

Polycrystalline samples of $Ca_3ReO_5Cl_2$ were synthesized by the conventional solid-state reaction. CaO, $ReO_3$, and $CaCl_2$ were mixed at a molar ratio of 2:1:1 in an argon-filled glove box, and the mixed powder was pressed into a pellet. The pellet was sealed in an evacuated quartz tube with being wrapped in a gold foil to prevent reaction with the quartz tube. The tube was heated at 800 °C for 24 hours. The sintered pellet was reground, pelletized, and heated again in the same way at 900 °C for 48 hours. The successful synthesis of a single phase of CROC was confirmed by powder x-ray diffraction (XRD, Rigaku RINT-2500) measurements using Cu-K$\alpha$ radiation: all the peaks in the XRD pattern were indexed to a orthorhombic unit cell with lattice parameters of $a$ = 11.892(3) Å, $b$ = 5.5606(12) Å, and $c$ = 11.129(4) Å, which are in good agreement with those reported for a single crystal of CROC [$Pnma$, $a$ = 11.8997(2) Å, $b$ = 5.5661(1) Å, $c$ = 11.1212(2) Å] [44].

Magnetic susceptibility and heat capacity measurements were conducted in a magnetic properties measurement system (MPMS-3, Quantum Design) and a physical properties measurement system (PPMS, Quantum Design), respectively. Heat capacity measurements down to 0.4 K were performed in the PPMS equipped with a $^3$He refrigerator.

First-principles calculations were performed based on the density functional theory (DFT), using a program package Quantum ESPRESSO [49] which employs plane-waves and pseudopotentials to describe the Kohn-Sham orbitals and the crystalline potential, respectively. The plane-wave cutoff for a wavefunction was set to 60 Ry. Calculations were performed with a GGA-PBE [50] functional using ultrasoft pseudopotentials [51]. We set $k$-point grids of Brillouin-zone integrations for the charge density to 5 × 10 × 5. Wannier functions were obtained by using a program package Wannier90 [52] which computes the maximally localized Wannier orbital.

## III. RESULTS

### 1 Crystal structure and magnetic interactions

The crystal structure of $Ca_3ReO_5Cl_2$ comprises $ReO_5$ square pyramids separated by $Ca_3Cl_2$ slabs. Figure 1(a) depicts the 3D network of the $ReO_5$ square pyramids which are not connected with each other by sharing common oxygen atoms. Considering that the distances between the $Re^{6+}$ ions along the $J$, $J'$, and $J''$ paths are 5.5661(1), 6.3989(3), and 5.5515(3) Å, respectively, the magnetic interactions seem to be weak and comparable to each other. However, the orbital state of the 5$d$ electron in the $Re^{6+}$ ion makes the magnetic interactions highly inequivalent. Previous optical measurements showed that one 5$d$ electron occupies the $d_{xy}$ orbital extending in the $bc$ plane, as illustrated in Fig. 1(b), where the $z$ axis is defined along the crystallographic $a$ axis and the $x$ and $y$ axes are toward the planer oxygen atoms along the crystallographic [011] and [0−11] directions, respectively. [44] One pair of the electron robes of the $d_{xy}$ orbital [yellow one in Fig. 1(b)] is aligned along the $b$ axis and overlaps with those of two neighboring $d_{xy}$ orbitals, which must cause a large direct exchange interaction to form an AF chain with $J$. On the other hand, another pairs of robes [blue one in Fig. 1(b)] are arranged in a staggered manner between the adjacent chains with a smaller overlap, resulting in the weak zigzag magnetic coupling of $J'$. In contrast, the out-of-plane exchange pass $J''$ must be much smaller because the $d_{xy}$ orbital lies in the $bc$ plane. Therefore, because of the specific arrangement of the $d_{xy}$ orbitals, the magnetic sublattice of CROC can be mapped to a quasi-2D ATL spin model.

### 2 DFT calculations

To confirm the implication mentioned above, we calculated the transfer integral $t$ and magnetic interaction $J$ (= 4$t^2/U$) by constructing the maximally localized Wannier orbital, based on first-principles DFT calculations [44]. The ratio of the magnetic interactions is obtained as $J : J' : J''$ = 1 : 0.295 : 0.0007. Note that the $J''$ is three orders of magnitude smaller than the leading interaction $J$ despite the shortest Re–Re distance, as expected from the arrangement of the $d_{xy}$ orbitals confined in the $bc$ plane. The in-plane anisotropy in the ATL of $J'/J$ = 0.295 is significantly large, which is comparable to ~0.3 for the representative ATL antiferromagnet $Cs_2CuCl_4$. [38–41] Therefore, CROC can be a good model compound for the ATL antiferromagnet.

### 3 Magnetic susceptibility

The temperature dependence of the magnetic susceptibility of a powder sample of $Ca_3ReO_5Cl_2$ (Fig. 2) exhibits a broad peak at around 28 K, which is characteristic of short-range order (SRO) for a low dimensional magnet. No anomaly indicative of a LRO is observed above 2 K. The magnetic susceptibility at high temperatures (150–350 K) is fitted to the Curie–Weiss equation including a temperature independent term $\chi_0$: $\chi = C/(T − \Theta_W) + \chi_0$, where $C$ is the Curie constant and $\Theta_W$ is the Weiss temperature. The fitting yields $C = 0.2958(2)$ cm$^3$ K mol$^{-1}$, $\Theta_W = −37.8(1)$ K, and $\chi_0 = −1.361(5)\times10^{-4}$ cm$^3$ mol$^{-1}$. The constant term $\chi_0$ must be mostly from the diamagnetism of core electrons, which is estimated to be $\chi_{dia} = −1.54\times10^{-4}$ cm$^3$ mol$^{-1}$ from the literature [53]. The inverse of the susceptibility after the subtraction of $\chi_0$ shows a linear temperature dependence, demonstrating well-defined Curie–Weiss behavior as shown in the inset of Fig. 2.

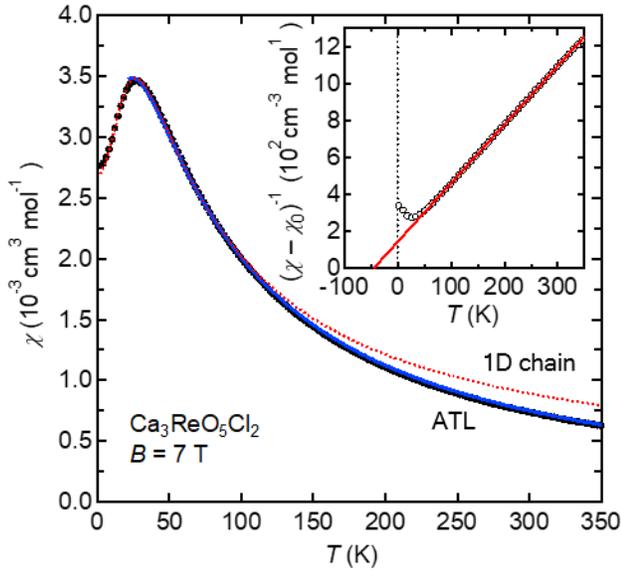

Fig. 2. Temperature dependence of magnetic susceptibility of a powder sample of Ca$_3$ReO$_5$Cl$_2$ (filled circles) measured upon cooling in a field of $B = 7$ T. The blue solid line shows a fit for $20 < T < 350$ K to the theoretical curve for the ATL model, which gives $J = 40.6$ K and $J'/J = 0.32$. The red dotted curve shows a fit for the low-temperature data between 2 and 70 K to the uniform spin-1/2 antiferromagnetic Heisenberg chain model, which yields $J = 41.3(1)$ K. The inset shows the inverse of the susceptibility after the subtraction of the temperature-independent term, $\chi_0 = −1.361\times10^{-4}$ cm$^3$ mol$^{-1}$, with a Curie–Weiss fit represented by the red solid line which gives $\Theta_W = −37.8(1)$ K.

The effective magnetic moment is calculated to be $\mu_{eff} = 1.585(2)$ $\mu_B$ from the Curie constant, which is significantly smaller than 1.73 $\mu_B$ expected for spin 1/2. This smaller effective moment is attributed to the reduction of $g$ factor from the spin-only value of 2 to 1.78 owing to the spin–orbit coupling (SOC); the SOC causes cancellation between spin and orbital moments for a less-than-half-filled $d$ shell. Because of the large SOC for the 5$d$ electron, the reduction is larger compared with those for the 3$d$ orbital; $g = 1.94$–1.98 for V$^{4+}$ ions with the 3$d^1$ electron configuration surrounded by oxygens in a square-pyramidal coordination [54]. Magnetic susceptibility measurements using a single crystal of CROC (not shown) reveal a small anisotropy, indicating that the isotropic Heisenberg spin model is an appropriate starting point: the $\chi$s along the $a$, $b$, and $c$ axes are scaled by the $g$ factors with 5% anisotropy. On the other hand, the considerably large, negative Weiss temperature of −37.8(1) K indicates significant antiferromagnetic interactions between the Re$^{6+}$ spins, even though the ReO$_5$ square-pyramids do not share their oxygens having the large distances more than 5.5 Å. This large magnetic interaction may reflect spatially extended 5$d$ orbitals mixed strongly with the surrounding oxygen 2$p$ orbitals.

As anticipated from the arrangement of the 5$d_{xy}$ orbitals and the DFT calculations, the temperature dependence of the magnetic susceptibility above 20 K is well reproduced by the high-temperature series expansion for the spin-1/2 Heisenberg ATL spin model [40], as shown in Fig. 2. The [5,5] Padé approximant [55] is used to extend the fitting range. The fitting yields $J = 40.6$ K and $J' = 13.0$ K ($J'/J = 0.32$) by employing the two parameters of $g = 1.78$ and $\chi_0 = −1.361\times10^{-4}$ cm$^3$ mol$^{-1}$ determined by the Curie–Weiss fitting. The obtained anisotropy reasonably agrees with $J'/J = 0.295$ from the DFT calculations. In the mean-field approximation for a paramagnetic phase in the high-temperature limit, all magnetic interactions additively contribute to a Weiss temperature as $\Theta_W = −[z\,S(S + 1)J_{av}]/3$, where $z$ is the number of neighboring ions and $J_{av}$ is the average interaction. Given that $J''$ is negligible, $\Theta_W = (2J + 4J')/4 = −33.3$ K, which is in good agreement with $\Theta_W = −37.8(1)$ K from the Curie–Weiss fitting. This consistency confirms that CROC is an ATL antiferromagnet with an intermediate anisotropy.

The low-temperature magnetic susceptibility of CROC below 70 K is well reproduced by the Bonner–Fisher curve for the spin-1/2 Heisenberg AF chain model [8]. The fit yields an intra-chain interaction $J = 41.3(1)$ K, $g = 1.573(5)$, and $\chi_0 = 1.69(12)\times10^{-4}$ cm$^3$ mol$^{-1}$; the effect of the interchain couplings seems to be incorporated into the reduced $g$ factor. Note that the $J$ value coincides with that from the fitting to the ATL model at high temperatures above the broad peak. Thus, to a first approximation, the low-temperature magnetic susceptibility of CROC is described in terms of quasi-1D chains with $J = 41$ K.

*4 Heat capacity*

Next, we show the heat capacity data down to 0.4 K [Fig. 3(a)]. A sharp peak indicative of a magnetic LRO of bulk nature is observed at $T_N = 1.13$ K. The $T_N$ is considerably reduced compared with $−\Theta_W = 37.8(1)$ K or $J = 40.6$ K owing to the low dimensionality and frustration. The peak slightly shifts to high temperature in an applied field of 5 T, which suggests that the magnetic ground state is not a simple AF order. A similar increase in transition temperature with the magnetic field is observed in the ATL magnet $Cs_2CuCl_4$, which is likely associated with an incommensurate (IC) helical order [56,57].

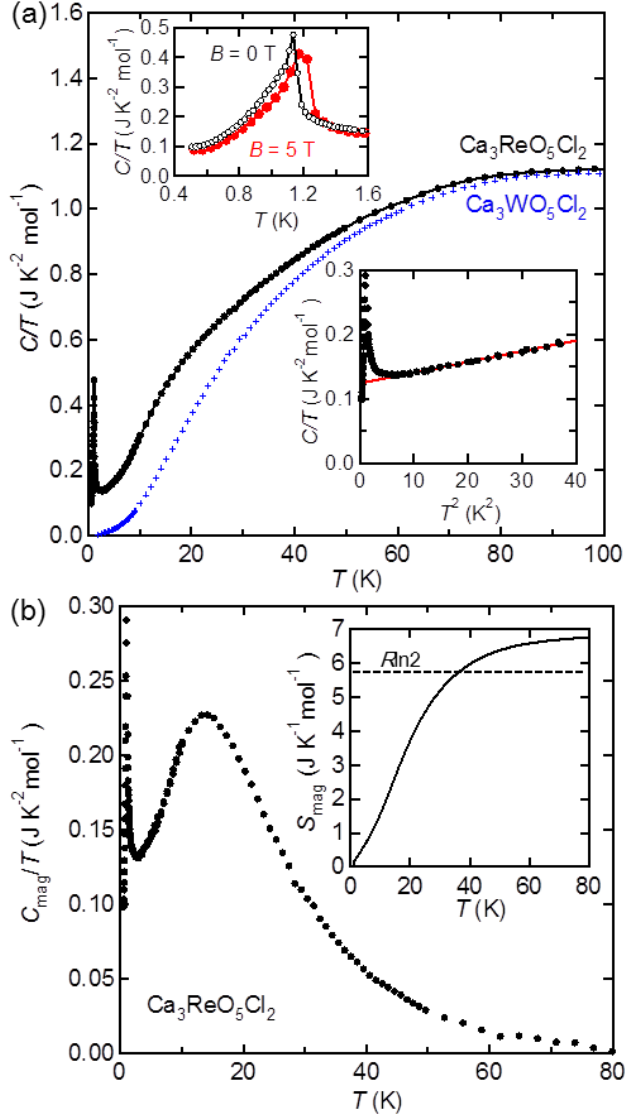

Fig. 3. (a) Heat capacity divided by temperature ($C/T$) for $Ca_3ReO_5Cl_2$ (black filled circles) and $Ca_3WO_5Cl_2$ (blue cross), the latter of which should give a reference to the common lattice contribution. The upper inset expands the low-temperature part around the peak at $T_N = 1.13$ K in fields of $B = 0$ (black open circles) and 5 T (red filled circles). The lower inset shows $C/T$ as a function of $T^2$ below 6 K with a fit to the equation $C(T)/T = \gamma + \beta T^2$ (red line). (b) Magnetic contribution to the heat capacity divided by temperature ($C_{mag}/T$) for $Ca_3ReO_5Cl_2$. The inset shows the magnetic entropy ($S_{mag}$).

A magnetic contribution to the heat capacity has been estimated by subtracting the heat capacity of the isostructural $d^0$ counterpart $Ca_3WO_5Cl_2$ from that of $Ca_3ReO_5Cl_2$, as shown in Fig. 3. The magnetic heat capacity divided by temperature $C_{mag}/T$ thus estimated shows a broad peak at around 14 K, as shown in Fig. 3(b), which must correspond to the development of SRO. For the spin-1/2 HAF chain it is known that a broad peak occurs at $J/3$ [7,8], which is 13.5 K for $J = 40.6$ K. The total magnetic entropy $S_{mag}$ over the entire temperature range reaches 6.74 J K$^{-1}$ mol$^{-1}$, which is slightly larger than the value of $R\ln 2$ = 5.76 J K$^{-1}$mol$^{-1}$ expected for spin 1/2, where $R$ is the gas constant; this deviation probably comes from experimental error underestimating the lattice contribution. The $S_{mag}$ released below $T_N$ is 0.16 J K$^{-1}$ mol$^{-1}$, which is only 2.7% of $R\ln 2$. An SRO that releases most magnetic entropy develops below ~20 K, followed by an LRO at a much-reduced temperature of $T_N = 1.13$ K.

An important finding here is the presence of a large $T$-linear heat capacity: the low-temperature heat capacity divided by temperature $C/T$ exhibits a linear dependence with a large intercept against $T^2$ in the temperature range between 2 and 6 K, as shown in the lower inset of Fig. 3(a). The intercept indicates a $T$-linear contribution to the heat capacity with a coefficient $\gamma = 114.7(4)$ mJ K$^{-2}$ mol$^{-1}$, which is obtained by fitting to the equation $C/T = \gamma + \beta T^2$, where the second term represents a lattice contribution in the low-temperature limit. Since CROC is an insulator, this $T$-linear term should not be ascribed to itinerant electrons but spins.

The $T$-linear heat capacity in the spin systems is known to arise exceptionally from a gapless spin excitation in the TLL in 1D spin-1/2 Heisenberg antiferromagnets such as $Cu(pyrazine)(NO_3)_2$ [58] and copper benzoate. [59] Moreover, similar $T$-linear heat capacity is found in a triangular lattice antiferromagnet having a spin liquid ground state in $\kappa$-$(BEDT-TTF)_2Cu_2(CN)_3$. [60] The $T$-linear heat capacity of CROC must originate from 1D spinon excitations in the ATL antiferromagnet. In the case of a spin-1/2 Heisenberg AF chain, the $T$-linear term is given as $C_{mag} = (2R/3) J \times T$ [7,8,61]. Applying this relation, we obtain $J(C) = 48.3(2)$ K, which reasonably agrees with the value estimated from the magnetization data; $J(\chi) = 40.6$ K. This agreement indicates that low energy excitation in CROC is characterized as that of TLL. Therefore, CROC behaves as a spin-1/2 Heisenberg AF chain at low temperatures, which is due to the geometrical frustration in the ATL antiferromagnet with a $J'/J$ ratio below 0.6 [24,25,28,30,31]; in fact, the experimentally estimated ratio is 0.32.

## IV. DISCUSSION

### 1 One-dimensionalization in the ATL antiferromagnet

It is now well established that although the interchain coupling is substantial in the ATL antiferromagnet, the frustration in the zigzag coupling markedly reduces interchain correlations in the ground state, which is called 'dimensional reduction' [24] or 'one-dimensionalization' [36]; dimensional reduction may be generally used, while one-dimensionalization is specific to the ATL antiferromagnet; very recently, Okuma et al. found a 3D spin system pharmacosiderite which exhibits a dimensional reduction to 2D, that is, two-dimensionalization by frustration. [62] As a result, the elementary excitations of the system are similar to those of 1D chains for the wide range of $J'/J \lesssim 0.6$.

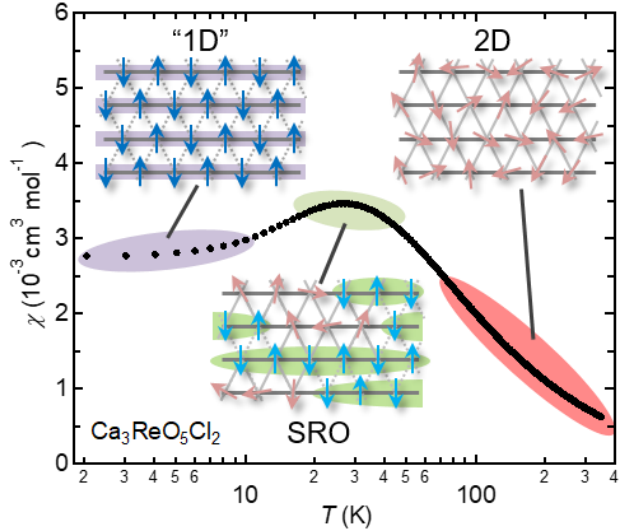

Fig. 4. Schematic representation of dimensional reduction or one-dimensionalization by geometrical frustration in the ATL antiferromagnet. The temperature evolution of the spin alignment is depicted in correspondence with the magnetic susceptibility of $Ca_3ReO_5Cl_2$. In a paramagnetic region at high temperatures, the system is regarded as a 2D ATL antiferromagnet with $J$ and $J'$. Upon cooling an AF correlation develops in every chain and induces 1D short-range order (SRO) at around the broad peak in the magnetic susceptibility. Eventually, as temperature approaches zero, growing 1D correlations effectively cancel out the interchain zigzag bonds (dotted lines), so that the system is considered to be a set of independent AF chains in the quantum critical regime.

Figure 4 illustrates the temperature evolution of magnetic correlations for the ATL antiferromagnet. At high temperatures in the Curie–Weiss regime, the system is a paramagnet having a 2D character with $J$ and $J'$, ignoring $J''$. Upon cooling, the magnetic susceptibility deviates from the Curie–Weiss behavior with the development of 1D SRO evidenced by the broad peak. Growing antiferromagnetic correlations in every chain effectively cancel out the interchain zigzag couplings between them, and, as the result, a dimensional crossover from 2D to 1D occurs; the cancellation is expected to be complete for Heisenberg interactions, while may be incomplete for higher-order couplings which must be much smaller. As the temperature approaches zero, the system is considered to consist of independent chains having infinite AF correlation lengths. This dimensional reduction by geometrical frustration is in contrast to the conventional dimensional crossover as a function of temperature for unfrustrated antiferromagnets: for example, in an anisotropic cubic antiferromagnet with $J \gg J' \gg J''$, the magnetic correlation changes upon cooling from quasi-1D to 2D and eventually to 3D.

Clear experimental evidence of dimensional reduction by frustration has been obtained in $Cs_2CuCl_4$. Coldea and coworkers carried out a series of INS experiments and observed a spinon continuum characteristic of the 1D chain [38,39]; in a quasi-1D system, deconfined spinons are formed as a domain wall between the two AF ground states. [24] However, the INS experiments reveal a significant dispersion in the excitation peaks along normal to the chain direction, indicating a substantial 2D character [39], which seems contradictory to the 1D approach. This fact is explained by introducing a 'triplon' which is a bound pair of 1D spinons: [35] whereas the 1D spinons are confined to the chains, the triplons can coherently move between chains, resulting in a lateral dispersion. It is noted that another theoretical approach based on a 2D spin liquid has been proposed for the interpretation of the experiments [63]; either of the theoretical approaches is quantitatively consistent with the INS results at $T \sim 0$, while there may be a difference in the temperature dependence of the dynamical spin structure factor.

In the present study, we show that $Ca_3ReO_5Cl_2$ is another good candidate to study the physics of the ATL antiferromagnet in detail. In spite of the 3D crystal structure, the specific arrangement of $d_{xy}$ orbitals in the $Re^{6+}$ ions gives a spin-1/2 ATL with $J = 41$ K and $J'/J = 0.32$ and with a negligible interplane coupling $J''/J = 0.0007$ from the DFT calculations. The one-dimensionalization by geometrical frustration is observed in the temperature evolution of magnetic susceptibility, and a gapless spin excitation of the 1D character is evidenced by the $T$-linear heat capacity with the coefficient scaled by $J$. Moreover, the peak temperatures associated with SRO in the magnetic susceptibility and heat capacity agree with those for spin-1/2 HAF chain. Future INS experiments would precisely determine the magnitude of the magnetic interactions and uncover the excitation spectrum of CROC which is to be compared with that of $Cs_2CuCl_4$.

## 2 Comparison with the other ATL antiferromagnets

Let us compare $Ca_3ReO_5Cl_2$ with the other ATL antiferromagnets, the characteristics of which are summarized in Table I. $Cs_2CuCl_4$ has $J = 4.3$ K and $J'/J \sim 0.34$ as determined by neutron scattering, or similar values from magnetic susceptibility and electron spin resonance (ESR) measurements [38–41]. The isostructural material

Cs$_2$CuBr$_4$ has $J$ = 15 K from the ESR measurements [41], and the $J'/J$ values are scattered as 0.467 [64–66] and 0.5–0.65 [67] from DFT calculations, 0.74 [66] by comparing the wave number of the IC helical order with the series expansion results, and 0.41 [41] from the ESR study. κ-(BEDT-TTF)$_2$Cu$_2$(CN)$_3$ has $J$ = 250 K [68] with the $J'/J$ values ranged between 0.3 and 1 [45–48]. Thus, $J$ = 41 K for CROC is between those of the inorganic and organic compounds, and the $J'/J \sim 0.3$ is comparable among them.

Table I. Model compounds of the ATL antiferromagnet. The leading exchange $J$, the $J'/J$ ratio, the LRO temperature $T_N$, and the frustration factor $f = J/T_N$ are compared.

| Compound | $J$ (K) | $J'/J$ | $T_N$ (K) | $f$ |
|---|---|---|---|---|
| Cs$_2$CuCl$_4$ [38–41] | 4.3 | 0.34 | 0.62 | 6.9 |
| Cs$_2$CuBr$_4$ [64–67] | 14.9 | 0.467–0.74 | 1.4 | 10.6 |
| κ-(BEDT-TTF)$_2$Cu$_2$(CN)$_3$ [45–48,68] | ~ 250 | 0.3–1 | < 0.032 | – |
| Ca$_3$ReO$_5$Cl$_2$ | 40.6 | 0.32 | 1.13 | 36 |

Compared with Cs$_2$CuCl$_4$, the $J$ value is almost one order larger, which means that the energy scale is much larger in CROC. This is experimentally advantageous to attain the low-temperature condition, $T \ll J/k_B$, while disadvantageous to attain the high-field condition, $H \gg J/k_B$, when studying the effects of magnetic field. The large $J$ value may be beneficial to achieving an ideal system with relatively small additional interactions such as farther-neighbor interactions or Dzyaloshinskii–Moriya (DM) interactions; for Cs$_2$CuCl$_4$ additional interactions of magnitude of only a few percents can induce entirely new phases in the phase diagram. [69] It is also pointed out that there is a wide $T$ range between $T_N$ and the peak temperatures in $\chi$ and $C_{mag}/T$, which are scaled by $J$. This makes it easy to study the $T$ dependence of the dynamical spin structure factor, which would give a further clue to the examination of the theoretical models. [63] Thus, CROC would provide us with complementary information in a wide parameter range and with a good test for theoretical approaches.

Cs$_2$CuCl$_4$ and Cs$_2$CuBr$_4$ undergo magnetic transitions to IC helical phases at $T_N$ = 0.62 K [56,57] and 1.4 K [64,65], respectively. The ratios of $J$ to $T_N$, defined as $f$, are 6.9 and 10.6, respectively. Compared with these values, $f$ = 36 for CROC is quite large. The magnetic order must be induced by finite interplane couplings $J''$ or anisotropy typically caused by DM interactions $D$. [70] Estimated for Cs$_2$CuCl$_4$ are $J'' = 0.045J$ and $D = 0.05J$ [39], which are large enough to induce an LRO. [63,69] The larger $f$ value of CROC indicates that these additional interactions are significantly smaller; this is partly due to the large $J$ value. In fact, the DFT calculation for CROC shows $J'' = 0.007J$. In addition, it is noted that the $J''$ bonds in CROC occur between neighboring chains halfway shifted to each other along the $b$ and $c$ axes, to generate a zigzag coupling [Fig. 1(a)], which may make the interplane coupling less effective, similar as for the zigzag coupling in the plane. Regarding the anisotropy, that in the $g$ factor is about 10% for Cs$_2$CuCl$_4$, while is smaller, 5%, for CROC. The magnitude of the DM interaction is not known for CROC but can be smaller. Therefore, CROC provides us with a good model system for studying one-dimensionalization by geometrical frustration in the ATL. The magnetic order of CROC remains to be examined by future neutron diffraction measurements, but may be a similar IC helical order, which is partly supported by the fact that the $T_N$ increases with increasing magnetic field [Fig. 3(a)]. [70]

Concerning the spin liquid state of κ-(BEDT-TTF)$_2$Cu$_2$(CN)$_3$, it was initially suggested to be an unconventional nonmagnetic Mott-insulating phase in the nearly isotropic triangular lattice [68]. Later, the first-principles calculations indicated a smaller ratio of 0.8 [45,47,48], and recent elaborate one mentioned even smaller values [46]. Thus, the observed gapless spin liquid may be a remnant of the TLL physics in a single chain [36,46]. If a translation into a Heisenberg model with infinite $U$ is acceptable, the compound realizes an ATL antiferromagnet in the family of organic compounds.

## V. SUMMARY AND REMARKS

We show that Ca$_3$ReO$_5$Cl$_2$ is a spin-1/2 ATL Heisenberg antiferromagnet with $J'/J$ = 0.32 and $J$ = 41 K, which is realized owing to the specific arrangement of the $d_{xy}$ orbitals of the Re$^{6+}$ ions. The observed $T$-linear heat capacity with the coefficient scaled by $J$ is due to a gapless excitation characteristic of a 1D spin system. This one-dimensionalization must be caused by geometrical frustration at the inter-chain zigzag couplings when AF magnetic correlations in every chain develop at low temperature. INS experiments are in progress to pin down the Hamiltonian of CROC and further bulk measurements under magnetic fields would reveal the magnetic phase diagram, which are to be compared in detail with those of Cs$_2$CuCl$_4$ and Cs$_2$CuBr$_4$.

From the materials point of view, there is a great advantage for this system: many related compounds having the general formula A$_3$BO$_5$X$_2$ (A = Ba, Sr, Ca, Pb; B = Re, W, Mo; X = Cl, Br, I) already exist and more wait to be synthesized in the family. We expect that tuning of the $J'/J$ ratio would be possible in A$_3$ReO$_5$X$_2$ depending on the sizes of the A and X ions, which enables us to test the general phase diagram of the ATL model. Thus, this group of compounds is potentially an excellent platform to explore quantum spin liquids, exotic magnetic phases, and novel magnetic excitations expected to occur in the ATL model.

## ACKNOWLEDGEMENTS

The authors thank M. Ogata for helpful discussions. Part of the numerical calculations in this work was performed using the supercomputer of the Institute for Solid State Physics, The University of Tokyo. This work was partly supported by Japan Society for the Promotion of Science (JSPS) KAKENHI Grant Number JP15K17695, JP18K13491 and by Core-to-Core Program (A) Advanced Research Networks. M. K and T. M are supported by Building of Consortia for the Development of Human Resources in Science and Technology from the MEXT of Japan.